# SpreadCluster: Recovering Versioned Spreadsheets through Similarity-Based Clustering

Liang Xu[1,2], Wensheng Dou[1*], Chushu Gao[1], Jie Wang[1,2], Jun Wei[1,2], Hua Zhong[1], Tao Huang[1]

[1]State Key Laboratory of Computer Science, Institute of Software, Chinese Academy of Sciences, Beijing, China
[2]University of Chinese Academy of Sciences, China
{xuliang12, wsdou, gaochushu, wangjie12, wj, zhonghua, tao}@otcaix.iscas.ac.cn

*Abstract*—**Version information plays an important role in spreadsheet understanding, maintaining and quality improving. However, end users rarely use version control tools to document spreadsheets' version information. Thus, the spreadsheets' version information is missing, and different versions of a spreadsheet coexist as individual and similar spreadsheets. Existing approaches try to recover spreadsheet version information through clustering these similar spreadsheets based on spreadsheet filenames or related email conversation. However, the applicability and accuracy of existing clustering approaches are limited due to the necessary information (e.g., filenames and email conversation) is usually missing.**

**We inspected the versioned spreadsheets in VEnron, which is extracted from the Enron Corporation. In VEnron, the different versions of a spreadsheet are clustered into an evolution group. We observed that the versioned spreadsheets in each evolution group exhibit certain common features (e.g., similar table headers and worksheet names). Based on this observation, we proposed an automatic clustering algorithm, *SpreadCluster*. SpreadCluster learns the criteria of features from the versioned spreadsheets in VEnron, and then automatically clusters spreadsheets with the similar features into the same evolution group. We applied SpreadCluster on all spreadsheets in the Enron corpus. The evaluation result shows that SpreadCluster could cluster spreadsheets with higher precision (78.5% vs. 59.8%) and recall rate (70.7% vs. 48.7%) than the filename-based approach used by VEnron. Based on the clustering result by SpreadCluster, we further created a new versioned spreadsheet corpus *VEnron2*, which is much bigger than VEnron (12,254 vs. 7,294 spreadsheets). We also applied SpreadCluster on the other two spreadsheet corpora FUSE and EUSES. The results show that SpreadCluster can cluster the versioned spreadsheets in these two corpora with high precision (91.0% and 79.8%).**

## I. INTRODUCTION

Spreadsheets are one of the most successful end-user programming platforms, and are widely used in various fields, such as finance, education, and so on [1]. Scaffidi [2] estimated that over 55 million end users in the United States worked with spreadsheets in 2012.

In conventional software development, source code can be managed by version control tools, e.g., SVN [3] and Git [4], and developers can reduce the cost and time by reusing or modifying existing code [5]. Similar to software development, end users may create new spreadsheets based on existing ones and reuse the data layout and computational logic (formulas). These new created spreadsheets share the same or similar data layout and computational logic with existing ones, and can be considered as the updated versions of the existing ones. Although there exist some version control tools for spreadsheets, such as SpreadGit [6] and SharePoint [7], spreadsheets are rarely maintained by these version control tools. The version information between spreadsheets is usually missing and different versions of a spreadsheet coexist as individual and similar spreadsheets. It is exhausting and time-consuming for end users to manage different versions of a spreadsheet, and it becomes more challenging when facing with a huge number of spreadsheets. For example, when users find that a spreadsheet contains an error, they need to manually identify all versions of this spreadsheet and recheck them, because they may contain the same errors. Recovering the version information will alleviate this situation. Further, the version information of spreadsheets can be used to study spreadsheet evolution [8][9], error and smell detection [10][11], and so on.

Existing approaches try to recover spreadsheet version information through clustering similar spreadsheets into *evolution groups*, based on the usage context of spreadsheets (e.g., spreadsheet filenames [8][10] and email conversation [10]). In this paper, we also use *evolution group* to denote a spreadsheet group whose spreadsheets are different versions of a spreadsheet. VEnron [8] clustered spreadsheets based on the similarity of spreadsheet filenames. Its basic idea is that different versions of a spreadsheet usually share the same shortened filenames after the version information (e.g., date, version number) in their filenames is removed. Users may share their spreadsheets to others through emails [12]. Schmitz et al. [10] found that the spreadsheets in the same email conversation may belong to the same evolution group, and further took the email conservation into consideration.

However, the applicability and accuracy of the filename-based [8] and email-conversation-based [10] spreadsheet clustering approaches are limited. The filename-based approach relies on the assumption that all spreadsheets are well-named. This assumption is not always true. First, no common practice is used for the naming of versioned spreadsheets. The different versions of a spreadsheet may have different filenames. The filename-based approach will cluster them into different evolution groups. Similarly, the spreadsheets with similar filenames may evolve from different spreadsheets and will be wrongly clustered together. Second, the filename-based clustering approach cannot cluster the spreadsheets whose filenames contain none or limited version information (e.g., 2003-01-36.xls). The email-conversation-based approach relies on the assumption that all spreadsheets are transferred by emails between users. However, the email conversations are not always available. First, collecting emails is difficult because they usually contain private information and users usually do not share their emails. Second, users may create their own spreadsheets and do not share them with anyone by emails. Third, users may send several completely different spreadsheets by an email, thus the spreadsheets in one email conversation may evolve from different spreadsheets. Therefore, a new spreadsheet clustering approach with higher applicability and accuracy will be appreciated.

---

* Corresponding author

**Table 1. Two evolution groups extracted from the VEnron corpus [8].**

| Group Name | Version Id | Spreadsheet Filename | Worksheet Name | | Subject of Related Email |
|---|---|---|---|---|---|
| 155_11_fomreq | v1 | May00_FOM_Req2.xls | May EPA Vols | FOM May Storage | Updated May '00 FOM requirements |
| | v2 | Jun00_FOM_Req.xls | Jun EPA Vols | FOM Jun Storage | CES FOM June '00 Requirements |
| | v6 | July00_FOM_Req.xls | Jul00 EPA Vols | FOM Jul Storage | CES FOM Volume Request for July 2000 |
| | v9 | Aug00_FOM_Req.xls | Aug00 EPA Vols | FOM Aug Storage | CES FOM August 2000 Volume request |
| 153_9_fom | e1 | FOM 0900.xls | Sept00 EPA Vols | FOM Sept Storage | September FOM volumes for CES  New Power |
| | e5 | FOM Oct-00.xls | Oct-00 EPA | October Storage | October 2000 FOM Requirements |
| | e8 | FOM Nov-00-1.xls | Nov-00 EPA | - | New Power November FOM - - Final Edition |
| | e9 | FOM Dec-00.xls | Dec-00 EPA | - | December 2000 FOM Estimates for New Power |

In this paper, we inspect the spreadsheets in each evolution group in VEnron [8], and observe that there are some similar features among the spreadsheets in each evolution group. For example, the spreadsheets in an evolution group share similar table headers and worksheet names. Based on this observation, we propose a novel spreadsheet clustering approach, named *SpreadCluster*, to identify evolution groups, which show spreadsheets are likely multiple versions evolved from the same spreadsheet. SpreadCluster first extracts these common features and calculates the similarity between spreadsheets based on these extracted features. Then, SpreadCluster uses the criteria about features learned from VEnron to cluster the spreadsheets into different evolution groups.

We compare SpreadCluster with the filename-based approach used in VEnron [8]. Our evaluation result shows that SpreadCluster obtains higher precision (78.5% vs. 59.8%), recall (70.7% vs. 48.7%) and F-Measure (74.4% vs. 53.7%) than the filename-based approach [8]. We further applied SpreadCluster on the other two big spreadsheet corpora, FUSE [13] and EUSES [14]. The evaluation results show that SpreadCluster can also achieve high precision (91.0% and 79.8%, respectively) on both corpus. Thus, SpreadCluster can be used to handle the spreadsheets in different domains Finally, based on the ground truth we built, we created a new versioned spreadsheet corpus *VEnron2*, which contains 1,609 evolution groups (12,254 spreadsheets). VEnron2 is much larger than its previous version VEnron (360 groups and 7,294 spreadsheets).

To our best knowledge, SpreadCluster is the first clustering approach that can automatically identify different versions of a spreadsheet by learning features and cluster them into an evolution group. The corpora we created are available online at http://www.tcse.cn/~wsdou/project/venron/.

In summary, this paper makes the following contributions:

- We propose SpreadCluster, a spreadsheet clustering approach that can automatically identify different versions of a spreadsheet with higher applicability and accuracy.

- We compare SpreadCluster with the filename-based approach used in VEnron [8]. Our evaluation result shows that SpreadCluster obtains higher precision, recall and F-Measure than the filename-based approach.

- We apply SpreadCluster on the other two big spreadsheet corpora, FUSE [13] and EUSES [14]. The evaluation results show SpreadCluster performs well in identifying different versions of spreadsheets used in different domains.

- Based on the ground truth we build, we further create a much larger versioned spreadsheet corpus from VEnron. Our new corpus VEnron2 is available online.

The remainder of this paper is organized as follows. Section II shows our motivation example and observations. Section III gives the detailed description of SpreadCluster. Section IV presents our evaluation. We discuss the issues and threats in Section V. Finally, we briefly introduce related work in Section VI, and conclude this paper in Section VII.

## II.  MOTIVATION

In this section, we illustrate two evolution groups extracted from VEnron [8]. We then introduce why need to cluster spreadsheets into evolution groups and why current approaches cannot work well through this example. Finally, we show the challenges in clustering the spreadsheets into evolution groups.

### A. Example

We take two evolution groups *155_11_fomreq* and *153_9_fom* in VEnron [8] as an example. These two evolution groups are used in the Enron Corporation [15]. The spreadsheets in both two groups are used to report the monthly and daily amount of "Baseload Storage Injections" in each month. More detailed information is shown in Table 1, including worksheet names, the subject of related emails, etc. We only show 4 (11 in total) spreadsheets in the group *155_11_fomreq* and 4 (9 in total) spreadsheets in the group *153_9_fom*, because the remaining spreadsheets are the same or similar to their previous versions. We only show two worksheet names ("-" means that the corresponding worksheets are absent). For other two worksheets in each spreadsheet, they have the same and fixed names ("Comments" and "Total Reqs"), as shown in Figure 1.

| | C | D | E | F |
|---|---|---|---|---|
| 1 | | | | |
| 4 | | June 2000 | | |
| 5 | Pipe/Service | Monthly | Daily | |
| 10 | Transco / AGL City Gate## | =E10*30 | 2411 | This volume included i |
| 11 | Sonat / ANR Shadyside | =E11*30 | 4285 | |
| 12 | Transco WSS | 62820 | =D12/30 | |
| 13 | Transco ESS | 3810 | =D13/30 | |
| 14 | Tennessee FS-MA | 0 | 0 | |
| 15 | Tennessee FS-PA | -12000 | =D15/30 | |
| 16 | CNG | 0 | 0 | |
| 17 | SONAT | 144036 | =D17/30 | |
| 18 | | | | |
| 19 | | | | |
| 20 | TCO FSS | 1089109 | =D20/30 | |
| 21 | | | | |

Comments | Total Reqs | Jun EPA Vols | FOM Jun Storage

(a) Jun00_FOM_Req.xls (v2)

| | C | D | E | F |
|---|---|---|---|---|
| 1 | | | | |
| 4 | | July 2000 | | |
| 5 | Pipe/Service | Monthly | Daily | This volume included i |
| 10 | Transco / AGL City Gate## | =E10*31 | 1694 | |
| 11 | Sonat / ANR Shadyside | =E11*31 | 2966 | |
| 12 | Transco WSS | 52700 | =D12/31 | |
| 13 | Transco ESS | 2666 | =D13/31 | |
| 14 | Tennessee FS-MA | =E14*31 | 0 | |
| 15 | Tennessee FS-PA | 0 | =D15/30 | |
| 16 | CNG | =E16*31 | 0 | |
| 17 | SONAT | 134001 | =D17/31 | |
| 18 | | =E18*31 | | |
| 19 | | =E19*31 | | |
| 20 | TCO FSS | 1089109 | =D20/31 | |
| 21 | | | | |

*The formula should be D15/31* (marked at cell E15)

Comments | Total Reqs | Jul00 EPA Vols | FOM Jul Storage

(b) July00_FOM_Req.xls (v6)

| | C | D | E | F |
|---|---|---|---|---|
| 1 | | | | |
| 4 | | September 2000 | | |
| 5 | Pipe/Service | Monthly | Daily | |
| 10 | Transco / AGL City Gate## | =E10*30 | | This volume included i |
| 11 | Sonat / ANR Shadyside | =E11*30 | | This volume was alread |
| 12 | Transco WSS | =E12*30 | 1741 | |
| 13 | Transco ESS | =E13*30 | 88 | |
| 14 | Tennessee FS-MA | =E14*30 | 0 | |
| 15 | Tennessee FS-PA | =E15*30 | 0 | |
| 16 | CNG | =E16*30 | 0 | |
| 17 | SONAT | =E17*30 | 4160 | |
| 18 | | | | |
| 19 | | | | |
| 20 | TCO FSS | =E20*30 | 36304 | |
| 21 | | | | |

Comments | Total Reqs | Sept00 EPA Vols | FOM Sept Storage

(c) FOM 0900.xls (e1)

Figure 1. Three real-world spreadsheets are extracted from two evolution groups in VEnron. The first two spreadsheets (a-b) come from *v2* and *v6* in group *155_11_fomreq*, and the third spreadsheet comes from *e1* in group *153_9_fom*. However, they can be considered part of one evolution group.

We only show three typical spreadsheets, as shown in Figure 1a-c. The first two spreadsheets, as shown in Figure 1a-b, are from the evolution group *155_11_fomreq*, and the last spreadsheet, as shown in Figure 1c, comes from the second evolution group *153_9_fom*. We can see that, all these spreadsheets share the similar semantics, and they should belong to the same evolution group. That said, these two evolution groups should be combined into one. More detailed information can be found in Section II.C.

### B. Why Should We Cluster Spreadsheets into Evolution Groups?

The version information among spreadsheets is usually missing, which makes it hard for end users to manage different versions of a spreadsheet. We outline two reasons why clustering spreadsheets into evolution groups can alleviate this situation.

#### 1) Easier to Find and Fix Spreadsheet Errors

Many techniques have been proposed to help developers to detect code clone and inconsistent errors by comparing multiple code clone fragments [16][17][18]. Similarly, we can find the inconsistent modifications on the spreadsheets by comparing two versions of a spreadsheet. These inconsistent modifications may indicate errors.

Figure 1a-b shows such a case. The worksheet *FOM Jun Storage* in Figure 1a shows the monthly and daily amount of storage injections in June, and the worksheet *FOM Jul Storage* in Figure 1b is an update for handling storage injections in July. We can see that they perform the same calculation, except the constants that are used in formulas. According to the table headers (Monthly and Daily), we can safely conclude that the constants (30 and 31) in the formulas are the numbers of the days in June and July. When the user created the spreadsheets for handling storage injections in July by reusing that in June, all constants in the formulas should change from 30 to 31. However, all the constants in the formulas are changed, except for that in the formula in cell E15 (marked by red rectangle). The formula has a potential error as users may enter non-zero value into cell D15.

To fix errors in the example, users could recheck all different versions of *July00_FOM_Req.xls* in Figure 1b. By clustering the different versions of a spreadsheet into an evolution group, users can cross-check them, and find opportunities on how to fix the errors. For example, users can find that the spreadsheet shown in Figure 1c gives a good example to fix the error.

#### 2) Easier to Understand Spreadsheet Evolution

After clustering the different versions of a spreadsheet into an evolution group, how bugs were introduced and fixed might be observed. For example, the formula error was introduced when the user created *July_FOM_Req.xls* in Figure 1b based on *Jun00_FOM_Req.xls* in Figure 1a. This formula error is hidden in the subsequent spreadsheets until the spreadsheet for September was created. Users may find that it is difficult to maintain these spreadsheets, and then refactored the spreadsheet. After that, the users used new filename naming pattern (e.g., *FOM 0900.xls* in Figure 1c) instead of the old one, in order to distinguish the refactored spreadsheets. We can see that the error is corrected and all formulas are placed in the column D.

### C. Existing Approaches

For the spreadsheets in Table 1, the filename-based approach used in VEnron [8] clusters these spreadsheets into two different evolution groups, because they have two different shortened names (*FOM_Req* and *FOM*, respectively). Since the email conservation is lost, some heuristic rules (e.g., the subjects of two emails should match after removing prefixes like "Re:", or the contents of one email exists in another one's.) are used to reconstruct the email conservation [10]. For example, as shown in Table 1, *v2* and *v9* are not in the same email conservation, due to the subjects do not match each other and the contents of emails are completely different. Thus, the email-conservation-based approach fails to cluster these spreadsheets, too.

However, the following evidences show that these spreadsheets of the two groups belong to the same evolution group.

- *Similar table headers and data layout*. As we can see from Figure 1, the worksheets of the spreadsheets in two

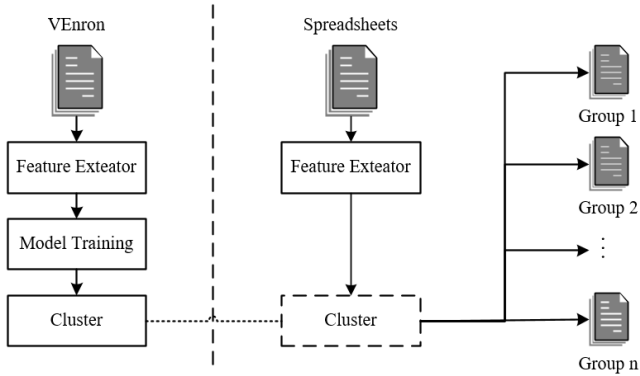

Figure 2. The overview of SpreadCluster.

groups share the same table headers, computational logic (formulas) and data layout.

- *Similar worksheets.* A worksheet is a function of the spreadsheet and the name usually indicates its main intent. As shown in Table 1, almost all spreadsheets of two groups contain four worksheets, only two spreadsheets contain the first three worksheets. Their names contain the same keywords that indicate the corresponding worksheets play the same roles in these spreadsheets.

- *The spreadsheets of two groups share one common maintenance staff.* The senders of the related emails can be considered as the maintenance staff of the spreadsheets. The spreadsheets in two groups are maintained by one common staff. Thus, it is likely that he/she created *FOM 0900.xls* in *153_9_fom* by reusing *Aug00_FOM_Req03.xls* in *155_11_fomreq*.

According to these evidences, we merged these two groups and form a bigger evolution group.

### D. Challenges and Approach Overview

To overcome the limitations of existing clustering approaches, we propose a novel, feature-based spreadsheet clustering approach. This approach calculates the similarity between spreadsheets using features and clusters the spreadsheets into different evolution groups.

There are several challenges in designing such a spreadsheet clustering approach. First, what features can be used to measure similarity between spreadsheets? Changes are common in spreadsheet reuse, not only the data values but also the data layout and computational logic. For example, end users may add/delete rows/columns, or add/delete/rename worksheets. The features selected should be as similar as possible within each evaluation group, and as different as possible from other groups. Second, how to define the similarity between spreadsheets? Compared with traditional software, spreadsheets have some special characteristics. For example, the data in spreadsheets is usually modified, two spreadsheets that have many differences on data with each other may still be different versions of the same spreadsheet. Existing spreadsheets comparison tools (e.g., SheetDiff [19] and xlCompare [20]) focus on finding and visualizing differences between two spreadsheets, and cannot be used to identify evolution groups. Third, how to determine the threshold for each feature? To determine the threshold, we need a training dataset. However, VEnron [8] cannot be used directly due to the drawback of the filename-based approach. For example in Figure 1, the versions of a spreadsheet are clustered into two independent groups.

To handle the first challenge, we manually inspected the evolution groups in VEnron. We observed that the spreadsheets in an evolution group usually share some common features. For example, as shown in Table 1, the corresponding worksheets share the same keywords in their names, and the tables in them share the same table headers. We discuss these features in Section III.A. To solve the second challenge, we take the characteristics of the spreadsheets into consideration. For example, the string in a cell is a complete information unit and we regard it as a word in the spreadsheet representation model that we use. More details can be found in Section III.C. The groups *155_11_fomreq* and *153_9_fom*, as shown in Table 1, are a good case that indicates how to handle the third challenge. Each group in VEnron is manually inspected by us to determine whether some groups should be merged into one according to several information (e.g., their contents, related email contents and whether they share common maintenance staffs). We give more details in Section IV.A.

## III. SPREADCLUSTER

SpreadCluster automatically identifies different versions of a spreadsheet. Figure 2 shows the overview of SpreadCluster. SpreadCluster contains two phases: a training phase and a working phase. In the training phase, SpreadCluster extracts features (Section III.A) from each spreadsheet (Section III.B). Then, SpreadCluster calculates the similarity between spreadsheets based on the extracted features (Section III.C). Finally, SpreadCluster trains a clustering model using the training dataset that is created based on VEnron [8]. In the working phase, SpreadCluster extracts features from spreadsheets and calculates the similarity between them. Then, SpreadCluster uses the trained model to cluster spreadsheets into different evolution groups (Section III.D).

### A. Feature Selection

In order to be broadly applicable, the features selected should exist in all spreadsheets, and reflect the spreadsheets' semantics. Although formulas are often used in spreadsheet analysis [19][20], we do not use formulas as a feature. It is because formulas may change, even after a simple row is added. In order to make our approach as simple as possible, we also tend to select a small number of features. We select two features as following:

*1) Table Header.* Table headers reflect the intended semantics of the processed data in a worksheet. Table headers have also been used to represent the possible semantics by existing work [11][23][24]. Because the spreadsheets in an evolution group share the same/similar semantics, the table headers in their spreadsheets are rarely changed. As shown in

Figure 3. A spreadsheet example for table header extraction.

Figure 1, the table headers (e.g., "Pipe/Service" and "SONAT") are the same in the three different versions.

*2) Worksheet Name.* Worksheet names denote the roles that the worksheets play in a spreadsheet. Generally, worksheet names give a high-level function description, and they are usually reused during evolution. For example, as shown in Figure 1a, the worksheet name *FOM Jun Storage* shows that the last worksheet of *v2* is used to store the data of Fom Storage in June. According to Table 1, we can see that the worksheet names only contain limited changes (e.g., from "FOM Jun Storage" in *v2* to "FOM Jul Storage" in *v6*).

### B. Feature Extraction

*1) Extracing Table Headers.* To extract table headers, we follow the region-based cell classification illustrated in UCheck [23]. The region-based cell classification first identifies the fences (a fence is a row or column that consists of blank cells, and could be the boundary of a table). After that, a worksheet is divided into one or more tables by the fences. In a table, row headers are usually located at rows in the top and column headers are usually located at columns in the left.

Unfortunately, the accuracy of this basic approach is not acceptable [24], because the spreadsheet allows users to design their table layout flexibly. Consider the example in Figure 3. This worksheet contains two tables (A1:E4 and A6:E9), and the second table A6:E9's headers are its first row (row 6). However, the user inserted one empty column in the second column of table A6:E9. The region-based cell classification approach identifies this worksheet as three small tables (A1:E4, A6:A9 and C6:E9) marked by red rectangles. Based on these three small tables, we can identify many data as headers, e.g., LIB in cell C7. Thus, we use the following three heuristics to avoid extracting the meaningless headers:

- *Row headers should occupy an entire row.* Not all tables have row headers, for example, the table A1:E4 in Figure 3. For this kind of tables, their top rows contain some cells with data value (e.g., number in the C1 and D1). Thus, we do not consider the strings in row 1 as row headers of table A1:E4.
- *Row/column headers cannot be a date or numeric sequence.* It is common that a table uses a date or numeric sequence as row/column headers, such as

"1,2, ..." and "2000/7/5, 2000/7/6, ...". This kind of table headers may appear in many unrelated spreadsheets, we avoid extracting this kind of table headers.

- *Row/column headers cannot be located in the right/bottom of a date or numeric sequence.* If some cells locate in the right/bottom of a date or numeric sequence, their contents are most likely to be data rather than table headers. For example, in Figure 3, C7:C9 are not the column headers of table C6:E9, since their left cells A7:A9 are a numeric sequence.

*2) Extracting Worksheet Names.* The worksheet name may contain version information (e.g., *FOM Jun Storage* in Figure 1a). Thus, we only extract meaningful words (e.g., *FOM Storage* in Figure 1a) from the worksheet name. These meaningful words usually indicate the function of the worksheet. To extract the keywords from the worksheet name, we remove stop words (e.g., "the" and "a"), special characters (e.g., "#" and "-"), spreadsheet related words (e.g., "Sheet"), number and date.

The length of a worksheet name is usually short, using the traditional method (e.g., edit distance) cannot assure the similar worksheet names have similar semantics. So we determine whether two worksheet names are similar according to whether they contain the same keywords. Since users may use default worksheet names (e.g., "Sheet1" and "Sheet2(1)") in different spreadsheets, we filter out the empty worksheets with default names from consideration.

### C. Similarity Measurement

In this section, we describe how to define the similarity between worksheets and that between spreadsheets.

*1) Similarity between Worksheets*

In each worksheet, table headers represent the semantics of the processed data. Thus, the similarity of two worksheets can be represented by the similarity of their table headers. We consider all table headers in a worksheet as a textual document, and then we can define the similarity of corresponding textual documents as the similarity of the worksheets.

The similarity between the textual documents have been well-studied in the area of information retrieval [25]. We select the widely-used Vector Space Mode (VSM) [26] as the document representation model. VSM converts every document into an $n$ dimension vector $<w_1, w_2, \cdots, w_n>$, where $n$ is the number of distinct words that exist in at least one document, and $w_i$ ($1 \le i \le n$) represents the degree of the importance of corresponding word to this document.

Since each table header should be taken as a whole to represent its semantics, we take each table header as a basic word. For example, in Figure 1a, the table header extracted from C5 is *Pipe/Service*. We treat this header as a word "Pipe/Service" instead of two words "Pipe" and "Service". Further, in order to avoid selecting useless words for clustering, we clean table headers by removing the meaningless words, including the common stop words (e.g., "all" and "above"), some spreadsheet related words (e.g., "#NAME?"), all special characters, strings of number/date, URL and mailing address. We also use the Porter Stemming

**Algorithm 1. Calculating similarity between spreadsheets**

**Input:** $sp_1, sp_2$ (two spreadsheets), $\theta_{ws}$ (threshold for the similarity between two worksheets)

**Output:** $S_{sp}$ (similarity score).

```
 1:  φ = ∅ ;   // Initialize the similar worksheet set
 2:  S_sp = 0;   // Initialize the similarity score
 3:  For each worksheet ws_i ∈ sp_1
 4:      For each worksheet ws_j ∈ sp_2
 5:          If S_ws(ws_i, ws_j) ≥ θ_ws and SimilarName(ws_i, ws_j)
 6:              φ = φ ∪ {ws_i, ws_j};   //Add two worksheets into φ
 7:          EndIf
 8:      EndFor
 9:  EndFor
10:  Calculate S_sp according to equation (4)
11:  Return S_sp
```

Algorithm [27], a widely used English stemming algorithm, to transform every word into its root form.

After the above steps, each worksheet is presented as a bag of words. We assign a weight to each word by utilizing the TF-IDF [28], whose value increases proportionally to the number of times a word appears in the worksheet, but is offset by the frequency of the word appears in other worksheets. Finally, the worksheets are presented as vectors. We give the formal representation of a worksheet as follows:

$$ws_i \equiv \overrightarrow{ws_i} \equiv (w_{i1}, w_{i2}, \cdots, w_{in}) \qquad (1)$$

Since a spreadsheet can be considered as a set of worksheets and each worksheet can be represented as a vector in VSM. Thus, we formally represent a spreadsheet as following:

$$SP \equiv \{ws_1, ws_2, \cdots, ws_k\} \qquad (2)$$

We use widely used cosine similarity to define the similarity between two worksheets as follows:

$$S_{ws}(ws_i, ws_j) = \frac{\overrightarrow{ws_i} \cdot \overrightarrow{ws_j}}{|\overrightarrow{ws_i}| \times |\overrightarrow{ws_j}|} \qquad (3)$$

- $ws_i$ and $ws_j$ are two worksheets.
- $\overrightarrow{ws_i}$ and $\overrightarrow{ws_j}$ are the vectors of $ws_i$ and $ws_j$, respectively.

Our extraction algorithm may fail to extract the headers and results in zero vector for a worksheet. (1) The worksheet is empty or only contains some types of data that cannot be handled by the Apache POI [29], e.g., charts. (2) The worksheet only contains data cells. For the above two cases, if their worksheet names share the same meaningful keywords, we set their similarity to 1, otherwise set to 0. Thus, $S_{ws}$ is within [0,1].

### 2) Similarity between Spreadsheets

Two spreadsheets with different numbers of worksheets may be different versions of the same spreadsheet. For example, as shown in Table 1, although the worksheet named *Storage* was deleted from *FOM Nov-00-1.xls*, it is still regarded as an update version of *FOM Oct-00.xls*, due to the fact that they share the main functions (three similar worksheets). Thus, the spreadsheet similarity should be able to tolerate the changes in the number of worksheets. We adapt

Jaccard similarity coefficient [30] to define the similarity between two spreadsheets. The Jaccard similarity coefficient is widely used to measure the similarity between finite sample sets. It is defined as the size of the intersection divided by the size of the union of the sample sets. The similarity between two spreadsheets is defined as follow:

$$S_{sp}(sp_i, sp_j) = \frac{|\varphi|}{|sp_i| + |sp_j|} \qquad (4)$$

- $|sp_i|$ and $|sp_j|$ are the numbers of the worksheets in the spreadsheets $sp_i$ and $sp_j$, respectively.
- $\varphi$ is a set of worksheets which come from the pairs $< ws_k, ws_l >$, where $ws_k \in sp_i$ and $ws_l \in sp_j$, and is bigger than the threshold $\theta_{ws}$, and their names share the same keywords. $|\varphi|$ is the number of different worksheets in $\varphi$. Note that $S_{sp}$ is within [0,1].

Algorithm 1 shows how SpreadCluster calculates the similarity between two spreadsheets $sp_1$ and $sp_2$. In order to get the set of worksheets $\varphi$, SpreadCluster calculates the similarity $S_{ws}(ws_i, ws_j)$ according to equation (3) for each worksheet $ws_i \in sp_1$ and each worksheet $ws_j \in sp_2$ (Lines 3-9). If $S_{ws}(ws_i, ws_j)$ is bigger than the threshold $\theta_{ws}$ and their names both contain the same keywords (Line 5), then we add $ws_i$ and $ws_j$ to the set $\varphi$ (Line 6). Finally, SpreadCluster calculates the similarity between two spreadsheets $S_{sp}$ according to equation (4).

We define the similarity between a spreadsheet $sp$ and a group $C$ as the maximum similarity achieved by $sp$ and spreadsheets in $C$ as follows:

$$S_{sc}(sp, C) = \max\left(S_{sp}(sp, sp_i)\right), sp_i \in C \qquad (5)$$

### D. Clustering Algorithm

Since users may choose the latest version to create new version every time, the accumulation of small changes may make the original version and the last version completely different. To handle this, we adapt the single-linkage algorithm [31] to cluster spreadsheets into evolution groups, as shown in Algorithm 2. First, we select a spreadsheet $sp_0$ from all spreadsheets $SP$ as a seed of group $C$ (Lines 3-6). Second, if there exists an un-clustered spreadsheet $sp$ and $S_{sc}$ ($sp$, $C$) is bigger than the threshold $\theta_{sp}$, then we add $sp$ into group $C$ and remove it from $SP$ (Lines 8-11), until no more spreadsheets can be clustered into $C$ (Lines 7-12). If group $C$ contains more than one spreadsheet, we consider that these spreadsheets are clustered successfully and assign a unique id for group $C$ (Lines 13-15). We repeat steps 1 and 2 until $SP$ is empty (Lines 2-16). Finally, our clustering algorithm returns all groups that contain more than one spreadsheet (Line 17).

### E. Threshold Learning

Two thresholds used by our clustering algorithm, $\theta_{ws}$ and $\theta_{sp}$, should be determined. They are used to determine whether two worksheets are similar and whether two spreadsheets belong to an evolution group, respectively.

**Algorithm 2. Clustering algorithm**

**Input:** $SP$ (all spreadsheets), $\theta_{WS}$ (threshold for the similarity between two worksheets), $\theta_{SP}$ (threshold for the similarity between two spreadsheets).

**Output:** $Cs$ (evolution group set).

```
1:   Cs = ∅ // Initialize group set
2:   While SP ≠ ∅
3:      C = ∅ // Initialize a new group
4:      Select a spreadsheet sp₀ ∈ SP
5:      SP = SP - {sp₀};  //Remove sp₀ from SP
6:      C = C ∪ {sp₀};  //Add sp₀ into group C
7:      Do
8:         If (∃ sp ∈ SP ∧ Ssc(sp,C) ≥ θsp)
9:            SP = SP - {sp};  //Remove sp from SP
10:           C = C ∪ {sp};  //Add sp into group
11:        EndIf
12:     While (C changes)
13:     If C contains more than one spreadsheet
14:        Cs = Cs ∪ C;  //Cluster successfully
15:     EndIf
16:  EndWhile
17:  Return Cs ;    //Return the clustering result
```

**Algorithm 3. Determine thresholds**

**Input:** $SP$ (training spreadsheet set)

**Output:** $\theta_{WS}$ (threshold for the similarity between two worksheets), $\theta_{SP}$ (threshold for the similarity between two spreadsheets).

```
1:   θWS = 0.01, θSP = 0.01  //Initialize
2:   While θWS ≤ 1
3:      While θSP ≤ 1
4:         groups=cluster SP by Algorithm 2 with (θWS, θSP);
5:         Calculate overall F-Measure for groups;
6:         Increase θSP by 0.01;
7:      EndWhile
8:      Increase θWS by 0.01;
9:   EndWhile
10:  Return θWS and θSP that achieve maximum F-Measure;
```

combination of $\theta_{ws}$ and $\theta_{sp}$. We choose the combination that can achieve the maximum F-Measure (Line 10).

## IV. EVALUATION

We evaluate SpreadCluster on three big spreadsheet corpora: Enron [12], EUSES [14] and FUSE [13]. We focus on the following research questions:

*RQ1 (Effectiveness): How effective is SpreadCluster in identifying different versions of spreadsheets? Specifically, what are the precision, recall and F-Measure?*

*RQ2 (Comparison): Can SpreadCluster outperform existing spreadsheet clustering techniques (e.g., the filename-based approach)?*

*RQ3 (Applicability): Can SpreadCluster cluster the spreadsheets from different domains?*

To answer RQ1, we evaluate SpreadCluster on the Enron corpus (Section IV.B.1). To answer RQ2, we compare SpreadCluster with the filename-based approach in terms of effectiveness on the Enron corpus (Section IV.B.2). To answer RQ3, we apply SpreadCluster on the EUSES [14] and FUSE [13] corpora (Section IV.B.3), and validate its precision. Our results are available online for future research (http://www.tcse.cn/~wsdou/project/venron/).

### A. Data Collection and Experimental Setup

We evaluate SpreadCluster on three widely used spreadsheets corpora: Enron [12], EUSES [14] and FUSE [13]. Enron is an industrial spreadsheets corpus, and contains more than 15,000 spreadsheets that were extracted from the Enron email archive [33]. EUSES is the most frequently used spreadsheet corpus, and contains 4,037 spreadsheets extracted from World Wide Web. FUSE is a reproducible, internet-scale corpus, and contains 249,376 unique spreadsheets that were extracted from over 26.83 billion webpages [34].

#### 1) Training Dataset based on VEnron

As discussed earlier in Section II.D, some groups in VEnron should be merged into one. Thus, VEnron cannot be used directly as training dataset. We manually inspected each evolution group in VEnron, and determined whether two groups should be merged into one according to the following criteria described in next section. Table 2 shows the final

Different combinations of values of $\theta_{ws}$ and $\theta_{sp}$ may result in different clustering results. We use the overall F-Measure [32] to measure how closely the clustering result $C = \{C_1, C_2, \cdots, C_m\}$ matches the manually clustering result $P = \{P_1, P_2, \cdots, P_n\}$. Here, $n$ and $m$ may not be equal.

Note that we do not know the correspondence between $P$ and $C$. To find the corresponding $C_i$ for each $P_j$, we first calculate precision, recall, and F-Measure for every $C_i$ and $P_j$ as follows:

$$precision(P_j, C_i) = \frac{|P_j \cap C_i|}{|C_i|} \qquad (6)$$

$$recall(P_j, C_i) = \frac{|P_j \cap C_i|}{|P_j|} \qquad (7)$$

$$F(P_j, C_i) = \frac{2 \times precision(P_j, C_i) \times recall(P_j, C_i)}{precision(P_j, C_i) + recall(P_j, C_i)} \qquad (8)$$

$$F(P_j) = \max_{i=1, 2, \ldots, m} F(P_j, C_i) \qquad (9)$$

For each $P_j$, $C_i$ that makes $F(P_j, C_i)$ to reach the maximum value is selected as the group corresponding to $P_j$. After getting all correspondence between $P$ and $C$, then the overall F-Measure is defined as follow:

$$F = \frac{\sum_{j=1}^{n} |P_j| \times F(P_j)}{\sum_{j=1}^{n} |P_j|} \qquad (10)$$

When the value of $F$ is closer to 1, the matching degree between the clustering result by our approach and the manually clustering result is higher.

Algorithm 3 shows how we learn the thresholds from a training dataset. Since the value range of $\theta_{ws}$ and $\theta_{sp}$ is between 0 and 1, we enumerate all possible combinations of $\theta_{ws}$ and $\theta_{sp}$, accurate to 0.01 (Lines 2-9), and then calculate the corresponding overall F-Measure (Line 5) for each training dataset. We merged 58 groups into 26 groups (Merged). We filtered out 6 groups which cannot be parsed by Apache POI [29] (Filter). Finally, we got 322 evolution groups containing 7,171 spreadsheets (TSet). We use this dataset to train SpreadCluster (Algorithm 3). In our experiment, we get $\theta_{WS} = 0.60$ and $\theta_{SP} = 0.33$.

*2) Validation Method*

Given a set of spreadsheets, we cluster them into different evolution groups using SpreadCluster or existing approaches. Since the creators of the spreadsheets used in our experiment are not available, we manually inspect each evolution group by ourselves. During our inspection, we use the spreadsheets' contents and associated information (e.g., emails), and try to answer the following questions and determine whether a spreadsheet belongs to an evolution group: (1) Are the spreadsheets in a group similar? (2) Do they share the same maintenance staffs? (3) Can we recover the order of these spreadsheets according to the time? We repeat the following steps until no further changes can be made.

i)    If all spreadsheets in a group are similar, we leave this group unchanged.

ii)   Otherwise, if we can find out some smaller groups whose spreadsheets are similar, we split the original group into subgroups, and each subgroup's spreadsheets are similar.

iii)  If only one spreadsheet is dissimilar to others in a group, we delete this spreadsheet from the group.

iv)   If any two spreadsheets are dissimilar in a group, we delete the group.

v)    If two groups are similar and can be merged, we merge them into one.

*3) Ground Truth*

In order to evaluate the recall of our approach, we need to obtain all evolution groups in Enron. However, the creators of Enron spreadsheets are not available, and we cannot obtain all these groups. We adopt a soft way to build the ground truth by combining all validated evolution groups by all approaches (SpreadCluster and the filename-based approach used in VEnron [8]). Note that some groups detected by two approaches will be merged if they contain common spreadsheets. We obtain 1,609 evolution groups, and 12,254 spreadsheets in total. We use these evolution groups as our ground truth.

For these 1,609 evolution groups, we further recover the order for the spreadsheets in each evolution group by following the order recovery rules in VEnron [8]. For example, in Figure 1, we can extract the date information (e.g., July and 0900) from the filenames, and determine the version order. After the version order recovery, we build a new versioned spreadsheet corpus VEnron2, which contains 1,609 evolution groups and 12,254 spreadsheets. VEnron2 is much larger than our previous versioned spreadsheet corpus VEnron [8] (7,294 spreadsheets and 360 groups).

*4) Evaluation Metrics*

Let $R_{clustered}$ denote the clustered groups and $R_{validated}$ denote the groups after manual validation. If a group in $R_{clustered}$ contains exactly the same spreadsheets that contained by a group in $R_{validated}$, we consider it correct (true positive).

**Table 2. The training dataset based on VEnron [8].**

|             | Original | Merged |    | Filter | TSet  |
|-------------|----------|--------|----|--------|-------|
| Groups      | 360      | 58     | 26 | 6      | 322   |
| Spreadsheets| 7,294    | 1,211  |    | 123    | 7,171 |

We define the *precision* of a clustering approach as the ratio of the number of groups clustered correctly to the number of groups in $R_{clustered}$, as shown below:

$$precision = \frac{|R_{clustered} \cap R_{validated}|}{|R_{clustered}|} \quad (11)$$

We use $R_{all}$ to denote all validated evolution groups in the ground truth. If a group in $R_{clustered}$ contains exactly the same spreadsheets that contained by a group in $R_{all}$, we consider it detected correctly. Thus, the *recall* and *F-Measure* are defined as follows:

$$recall = \frac{|R_{clustered} \cap R_{all}|}{|R_{all}|} \quad (12)$$

$$F\text{-}Measure = \frac{2 \times precision \times recall}{precision + recall} \quad (13)$$

**B. Experimental Results**

*1) RQ1: Effectiveness*

We apply SpreadCluster on all the spreadsheets in Enron, and further manually validate all groups in the clustering result. Table 3 shows the detected and validated results. We can see that SpreadCluster clusters all spreadsheets into 1,561 groups (Detected). Among these groups, 1,226 are correctly clustered (Correct). The precision of SpreadCluster is 78.5%, which is promising.

As shown in Table 4, among 1,609 evolution groups in the ground truth (GroundTruth), SpreadCluster can detect 1,137 evolution groups correctly (Correct). Note that these 1,137 evolution groups are less than validated groups detected by SpreadCluster (1,137 vs. 1,226). It is because in the ground truth, some groups merge the groups detected by two approaches and they may contain spreadsheets that cannot be detected by SpreadCluster. Thus, the recall and F-measure of SpreadCluster are 70.7% and 74.4%, respectively.

Therefore, we can draw the following conclusion:

> SpreadCluster can identify evolution groups with high precision (78.5%) and recall (70.7%).

We further investigate why SpreadCluster fails to detect some evolution groups. First, although some spreadsheets share common or similar worksheets, we do not think that they are different versions of the same spreadsheet and cluster them into different evolution groups. This is the main reason we need to delete/split some groups or delete some spreadsheets, as shown in Table 3. Second, some spreadsheets only contain data, charts or numeric/date sequence as table headers, SpreadCluster cannot detect table headers in them. Specially, some spreadsheets share some empty worksheets with same names. Third, our header extraction algorithm is heuristic and it may fail in some cases. We will further improve the header extraction algorithm in the future.

**Table 3. The clustering results of SpreadCluster and the filename-based approach on three spreadsheet corpora.** For each corpus, columns 3-11 show the numbers of evolution groups. After SpreadCluster (or the filename-based approach [8]) detected evolution groups (Detected), we manually validated all or parts of them (Validated). We confirmed some of them are correct (Correct), and deleted groups when all their spreadsheets are dissimilar (Deleted), and deleted some spreadsheets if they are different from others (DeleteSpread). Further, if a group contains more subgroups, we split it into more groups (Split). Some groups were merged into other groups (Merged). We may perform deleting spreadsheets, merging groups, or splitting groups together (MultiOp). Column 11 (Final) shows our validated results.

| Approaches | Corpus | Detected | Validated | Correct | Deleted | DeleteSpread | Split | Merged | MultiOp | Final | Precision |
|---|---|---|---|---|---|---|---|---|---|---|---|
| SpreadCluster | Enron | 1,561 | 1,561 | 1,226 | 59 | 69 | 33 | 125 | 49 | 1,507 | 78.5% |
|  | EUSES | 213 | 213 | 170 | 36 | 7 | - | - | - | 177 | 79.8% |
|  | FUSE | 10,985 | 200 | 182 | 6 | 1 | 2 | 9 | - | 188 | 91.0% |
| Filename-based [8] | Enron | 1,613 | 1,613 | 965 | 153 | 88 | 15 | 341 | 51 | 1,278 | 59.8% |

**Table 4. The comparison of SpreadCluster and the filename-based approach on Enron.** For each group in ground truth (GroundTruth), we validated whether it is detected correctly by two approaches. We confirmed some groups are correctly detected (Correct), none spreadsheets in some groups are clustered (GMissed), some groups are incomplete (FMissed), some groups are clustered into several small subgroups (Split), some groups are mixed with other groups (Mixed), some groups are involved in more than one case in the above (MultiCase), for example, one group is split into subgroups and some spreadsheets in it are missed.

| Approach | GroundTruth | Correct | GMissed | FMissed | Split | Mixed | MultiCase | Recall | F-Measure |
|---|---|---|---|---|---|---|---|---|---|
| SpreadCluster | 1,609 | 1,137 | 90 | 91 | 37 | 223 | 31 | 70.7% | 74.4% |
| Filename_based [8] |  | 783 | 272 | 164 | 92 | 217 | 81 | 48.7% | 53.7% |

### 2) RQ2: Comparison

We further compare the effectiveness of SpreadCluster with the filename-based approach [8]. We do not compare with the email-conversation-based approach [10] because it is challenging to automatically reconstruct email conservation precisely. We choose Enron [12] to evaluate these two approaches rather than EUSES [14] and FUSE [13]. First, the spreadsheets in EUSES are usually independent. Second, the filename-based approach cannot work on FUSE because the spreadsheets in FUSE were all renamed as a combination of numbers and letters with fixed length of 36 (e.g., "00001ca0-d715-4250-bba8-f416281ffb1c").

Table 3 also shows the detected and validated clustering results of the filename-based approach on Enron. We can see that the filename-based approach, among 1,613 detected groups (Detected), 956 groups are correctly detected (Correct). The precision of filename-based approach is 59.8%, which is much lower than SpreadCluster (78.5%).

We further compared the recall and F-Measure. From Table 4, we can see that the filename-based approach can only detect 783 evolution groups correctly (Correct). Thus, its recall and F-measure are 48.7% and 53.7%, which are also much lower than SpreadCluster (70.7% and 74.4%).

Therefore, we draw the following conclusion:

> SpreadCluster performs better than the filename-based approach in identifying evolution groups.

We further compare the two approaches in more details to understand why SpreadCluster performs better. The detailed result is shown in Table 4. SpreadCluster misses much less evolution groups than the filename-based approach (GMissed; 90 vs. 272). This indicates that spreadsheets in many evolution groups do not have similar filenames (as assumed in the filename-based approach [8]), and thus the filename-based approach would fail to detect them. Further, SpreadCluster can detect evolution groups more precisely, e.g., evolution groups that were split (Split) and incomplete groups (FMissed) are much lesser, too. Thus, for the filename-based approach, its accuracy and applicability heavily depend on the spreadsheet filenames. SpreadCluster can overcome this limitation and achieves higher accuracy.

### 3) RQ3: Applicability

The spreadsheets in the Enron dataset were created to store or process the data in the financial area, they are domain-specific. To validate whether SpreadCluster can identify evolution groups from other domains, we apply SpreadCluster on the FUSE [13] and EUSES [14] corpora. These two corpora were extracted from the web pages and used for different domains. Since there is no training dataset can be used to learn the thresholds for FUSE and EUSES. we apply SpreadCluster on these two corpora with the thresholds trained from VEnron [8].

Table 3 shows the detected results. SpreadCluster can detect 10,985 groups (Detected) on FUSE. It is impractical to validate all these groups manually, thus we randomly selected some groups to validate and estimated the accuracy of SpreadCluster on FUSE. In order to alleviate human labor, we randomly selected 200 groups containing no more than 20 spreadsheets to validate, since only 279 groups contain more than 20 spreadsheets. In Table 3, we can see that SpreadCluster can achieve 91.0% precision on FUSE [13], which is higher than Enron (78.5%).

We further applied SpreadCluster on the EUSES [14] corpus to find the hidden different versions of spreadsheets in EUSES. SpreadCluster clustered only 481 of 4,140 spreadsheets into 213 groups. It is not surprised, since many versions of a spreadsheet had been have been cleaned as duplicated spreadsheets in EUSES. We manually validated all groups since the number of groups is not large. As shown in Table 3, the precision of SpreadCluster achieves 79.8% (170 of 213), which is a little lower than that on FUSE, but still higher than that on Enron.

Therefore, we draw the following conclusion:

> SpreadCluster performs well in identifying evolution groups for different spreadsheet corpora in different domains.

## V. Discussion

While our experiments show that SpreadCluster is promising, we discuss some potential threats to our approach.

***Representativeness of our experimental subjects.*** One threat to the external validity is the representativeness of experimental subjects used in our evaluation. We select the Enron [12], EUSES [14] and FUSE [13] that are the three biggest spreadsheet corpora so far, and have been widely used for spreadsheet-related studies [10][35][36][37].

***Training dataset and evolution group validation.*** To construct the training dataset and validate the clustering results, we manually inspected spreadsheets in each evolution group. However, we cannot guarantee that this dataset does not contain any false positives or false negatives. To minimize this threat, the groups were cross checked by two authors.

***Ground truth used in the experiments.*** Since it is impractical to obtain all evolution groups in Enron, we build the ground truth by combining the validated results of two approaches (SpreadCluster and the filename-based approach [8]). This ground truth may contain some biases although we have done our best to avoid that. In the future, we will try to get a complete ground truth in a small corpus.

***Similarity definition and clustering algorithm.*** The similarity definitions and clustering algorithm we used is simple and effective regarding to our presentation model. Different definitions and clustering algorithm may achieve better results, and we will explore that in the future.

***Parallel evolution.*** Spreadsheets can be forked like software and evolve in parallel. Our approach clusters the spreadsheets in parallel evolution groups into the same group. This problem may be solved by using more information (e.g., spreadsheet filenames). We leave this as future work.

## VI. Related Work

We focus on these pieces of work concerning spreadsheet corpora, clone detection, evolution and error detection.

***Spreadsheet corpora.*** EUSES [14] is the most widely spreadsheet corpus, containing 4,037 spreadsheets. Enron [12] is the first industrial spreadsheet corpus, containing more than 15,000 spreadsheets extracted from the Enron email archive [33]. FUSE [13] is the biggest spreadsheet corpus, containing 249,376 spreadsheets extracted from over 26 billion pages [34]. The spreadsheets in these three corpora are independent and all relationships between them were missing. SpreadCluster can recover the relationships between spreadsheets by detecting evolution groups. VEnron [8] is the first versioned spreadsheet corpus, containing 360 evolution groups and 7,294 spreadsheets. VEnron uses the filename-based approach to identify evolution groups, and leads to inaccurate and incomplete results. We applied SpreadCluster to create a much bigger versioned corpus than VEnron.

***Spreadsheet clone detection.*** Hermans et al. [38] proposed data clone detection in spreadsheets. TableCheck [11] identifies table clones that share the same/similar computational semantics. These two approaches can only identify areas with the same data or computational semantics. However, changes (e.g., new data, formula) are common in spreadsheets, clone detection techniques cannot be employed to identify evolution groups. Spreadsheet comparison tools, like SheetDiff [19] and xlCompare [20], can be used to find differences between spreadsheets. However, they cannot judge whether two spreadsheets belong to an evolution group.

***Spreadsheet evolution.*** Due to the version information is usually missing, few work focus on spreadsheet evolution. Hermans et al. carry out an evolution study on 54 pairs of spreadsheets [9]. The spreadsheet evolutionary characteristics (e.g., the level of coupling) were observed by comparing each pair of spreadsheets. But the studied spreadsheets are not publicly available. Dou et al. [8] study spreadsheet changes from multiple views (e.g., formula, entered value and error trend) during evolution. SpreadCluster's results can be used to do further spreadsheet evolution studies.

***Spreadsheet error detection.*** Various techniques have been proposed to detect spreadsheet errors. UCheck [23] and dimension check [39] infer the types for cells and use a type system to carry out inconsistency checking. Dou et al [40][41] extract cell arrays that share the same computational semantics, then find and repair inconsistent formulas and data by inferring their formula patterns. Hermans et al. [32][33] adjust and apply code smells on spreadsheets. CheckCell [43] detects data value that affects the computation dramatically. However, these pieces of work focus on a single spreadsheet. SpreadCluster makes it possible to detect errors or smells caused by inconsistent modifications in spreadsheets by comparing different versions of a spreadsheet.

## VII. Conclusion and Future Work

In this paper, we propose SpreadCluster, a novel clustering algorithm that can automatically identify different versions of a spreadsheet. SpreadCluster calculates the similarity based on the features in spreadsheets, and clusters them into evolution groups. Our experimental result shows that SpreadCluster can improve the filename-based clustering approach greatly. We also apply SpreadCluster on FUSE [13] and EUSES [14], and it can also achieve high precision. That indicates SpreadCluster can perform well in identifying evolution groups in different domains.

We further build a new versioned spreadsheet corpus based on the ground truth we used, VEnron2, which contains 1,609 evolution groups and 12,254 spreadsheets. VEnron2 is much larger than our previous versioned spreadsheet corpus VEnron [8] (360 groups and 7,294 spreadsheets). Our new corpus VEnron2 is now available online for future research (http://www.tcse.cn/~wsdou/project/venron/).

We plan to pursue our future work in three ways. (1) For now, we manually recover the version order among spreadsheets. We will study how to automatically recover the version order. (2) SpreadCluster can be further improved by more precise header extraction algorithm. (3) An empirical study on the versioned spreadsheets can be conducted to improve the understanding of spreadsheet evolution.

## Acknowledgment

This work was supported in part by Beijing Natural Science Foundation (4164104), National Key Research and Development Plan (2016YFB1000803), and National Natural Science Foundation of China (61672506).